# Analysis of the Evolution of Hydrodynamic Instability at the Interface of Active and Passive Media

S.E. Kuratov, A.Yu. Mikulin

(FSUE VNIIA, Moscow, Russia)

The evolution of an instability at the interface of active and passive media is considered. An asymptotic form of a collision integral is found and the limitations of hydrodynamic approach are determined. A growth increment of small perturbations of the interface is found using the potential flows theory. The obtained results were applied to estimation of the instability contribution to the development of some known physical processes.

## Introduction

The instability phenomenon considered in this paper consists in the increase of perturbations at the interface of active and passive media. The active medium is a substance in which a reaction is accompanied by the generation of high-energy particles. These particles decelerated in the both media which produce a force operating in the region near the interface.

The hydrodynamic approach was applied to analyze this instability. The validity conditions of hydrodynamic approach are determined by means of the Chapman-Enskog method.

We present a list of known physical processes where the evolution of the discussed instability could be essential, and the contribution to each process is estimated.

## Justification of Hydrodynamic Approximation

According to the kinetic theory [1], it is the Maxwell distribution function that corresponds to the equilibrium state when the Boltzmann collision integral vanishes. Due to the interactions with the fast products of the reaction the distribution function of the particles of the passive medium is distorted and becomes non-maxwellian. Let us estimate the correction term to the distribution function in the framework of the Chapman-Enskog method.

We consider the most "non-equilibrium" case: the system consists of a light passive medium and a heavy active medium where fast and heavy products of the reaction are generated.

The Boltzmann equation for light particle distribution looks as follows

$$\frac{\partial f_L}{\partial t} + \vec{\upsilon}_L \frac{\partial f_L}{\partial \vec{r}} + \vec{F}_L \frac{\partial f_L}{\partial \vec{p}} = I_{st}(f_L, f_L) + I_{st}(f_L, f_A) + I_{st}(f_L, f_H) \quad (1)$$

where $f_L$ - light particle distribution function, $f_H$ - heavy particle distribution function, $f_A$ - distribution function of reaction products, $I_{st}(f_a, f_b) = \iint d\vec{p}_a d\sigma_{ab} \upsilon_{ab} \{f'_a f'_a - f_a f_b\}$ - Boltzmann collision integral.

We exclude $I_{st}(f_L, f_H)$ since at this stage we neglect the diffusion and assume that there are no active medium particles inside the passive medium. The equation (1) will take the following form:

$$\frac{\partial f_L}{\partial t} + \vec{v}_L \frac{\partial f_L}{\partial \vec{r}} + \vec{F}_L \frac{\partial f_L}{\partial \vec{p}} = \left[ I_{st}(f_L, f_L) + I_{st}(f_L, f_A) \right]$$

Let us apply the Chapman-Enskog method [2].

For distributions which differ slightly from the equilibrium ones a distribution function can be represented as a functional series:

$$f_L = f_L^{[0]} + f_L^{[1]} + f_L^{[2]} + ...$$

Here we imply that $f_L^{[0]} \gg f_L^{[1]} \gg f_L^{[2]} \gg ...$

We represent the first order correction term for the light particle distribution function in the form

$$f^{[1]} = f^{[0]} \varphi, \quad \varphi \ll 1$$

In accordance with Ref. [2] we obtain

$$f_L^{[0]} \left\{ \vec{V}_L \frac{\partial \ln T_L}{\partial \vec{r}} \left[ \frac{m_L V_L^2}{2kT_L} - \frac{5}{2} \right] + \frac{m_L}{kT_L} \left( V_{L,i} V_{L,k} - \frac{1}{3} \delta_{ik} V_L^2 \right) \frac{\partial v_{0,L,i}}{\partial r_k} + \frac{1}{kT_L} (\vec{V}_L, \vec{F}) \right\} =$$
$$= \int d\vec{p}_{L,1} |\vec{V}_L - \vec{V}_{L,1}| d\sigma f_L^{[0]}(\vec{V}_L) f_L^{[0]}(\vec{V}_{L,1}) \left[ \varphi(\vec{V}_L') + \varphi(\vec{V}_{L,1}') - \varphi(\vec{V}_L) - \varphi(\vec{V}_{L,1}) \right] + \quad (2)$$
$$+ I_{st}\left( f_L^{[0]}, f_A \right)$$

Deducing the left hand side of Eq. (2) we take into account the appearance of force

$$\vec{F} = \frac{1}{n_L} \int I_{st}\left( f_L^{[0]}, f_A \right) m_L (\vec{v}_L - \vec{v}_{0,L}) d\vec{p}_L$$

in the hydrodynamic equations.

Let us represent $\varphi$ as a sum of $\varphi_1, \varphi_2$ and $\varphi_3$. The functions $\varphi_2$ and $\varphi_3$ are the known solutions, namely, corrections to the light particle distribution function caused by the inhomogenity of their temperatures and the average mass velocity. The sum of solutions $\varphi_2$ and $\varphi_3$ satisfies the Eq. (2) without the terms $I_{st}\left(f_L^{[0]}, f_A\right)$ and $\dfrac{f_L^{[0]}}{kT_L}\left(\vec{V}_L, \vec{F}\right)$. Hence the following equation determines $\varphi_1$:

$$\int d\vec{p}_{L,1}\left|\vec{V}_L - \vec{V}_{L,1}\right| d\sigma f_L^{[0]}\left(\vec{V}_L\right) f_L^{[0]}\left(\vec{V}_{L,1}\right)\left[\varphi_1\left(\vec{V}_L^{'}\right) + \varphi_1\left(\vec{V}_{L,1}^{'}\right) - \varphi_1\left(\vec{V}_L\right) - \varphi_1\left(\vec{V}_{L,1}\right)\right] + \\ + I_{st}\left(f_L^{[0]}, f_A\right) = 0 \qquad (3)$$

Now we determine the collision integral $I_{st}\left(f_L^{[0]}, f_A\right)$. Let us consider that the plane layer penetrates into the light medium. The velocities of all particles are equal in the absolute value and direction. We assume the time dependence of the momentum of reaction products to be known (the so called approximation of continuous deceleration [3]), and turn to the reference frame associated with them. Then the collision of the heavy product of the reaction with the light particle does not change the pulse absolute value of the latter. In this reference frame the expression for the Maxwell distribution function of light particle has the form:

$$f_L^{[0]}\left(\vec{p}_L\right) = \dfrac{n_L}{(2\pi m_L kT_L)^{\frac{3}{2}}} \exp\left(-\dfrac{\left(\vec{p}_L + m_L \vec{\upsilon}_A(t)\right)^2}{2m_L kT_L}\right)$$

$\vec{\upsilon}_A(t)$ - reaction product velocity in the laboratory reference frame, $\vec{p}_L$ - light particle momentum in the center-of-mass system.

For scattered light particles the distribution function takes the following form

$$f_L^{[0]}\left(\vec{p}_L^{'}\right) = \dfrac{n_L}{(2\pi m_L kT_L)^{\frac{3}{2}}} \exp\left(-\dfrac{\left(\vec{p}_L^{'} + m_L \vec{\upsilon}_A(t)\right)^2}{2m_L kT_L}\right)$$

where $\vec{p}_L^{'} = \mu_{LA}\upsilon_{LA}\vec{n}^{'} + \dfrac{m_L}{m_L + m_A}\left(\vec{p}_L + \vec{p}_A\right)$

In the limit $m_A \gg m_L$ is $\vec{p}_L^{'} = p_L \vec{n}^{'}$

In the reference frame the reaction products are at rest. So $\int d\vec{p}_A f_A = \int d\vec{p}_A f_A^{'} = n_A$. If we assume that the scattering cross-section is isotropic and does not depend on the relative velocity of colliding particles, the Boltzmann collision integral takes the form

$$I_{st}\left(f_L^{[0]}, f_A\right) = n_A \upsilon_L \sigma_{AL} \int \frac{d\mathrm{O}_{\vec{n}'}}{4\pi} \left(f_L^{[0]}(\vec{p}_L') - f_L^{[0]}(\vec{p}_L)\right)$$

The calculations result in:

$$I_{st} = \frac{n_L n_A \sigma_{AL}}{(2\pi m_L kT_L)^{\frac{3}{2}}} \exp\left(-\frac{p_{L,C}^2 + m_L^2 \upsilon_A^2(t)}{2m_L kT_L}\right) \cdot$$
$$\cdot \left[\frac{kT_L}{m_L \upsilon_A(t)} \sinh\left(\frac{p_{L,C} \upsilon_A(t)}{kT_L}\right) - \frac{p_{L,C}}{m_L} \exp\left(-\frac{p_{L,C} \upsilon_A(t) \cos \Psi_C}{kT_L}\right)\right] \quad (4)$$

where $\Psi_C$ - angle between the velocities of a light particle and a reaction product

We can easily pass from this expression to the collision integral in the laboratory reference frame by a substitution in Eq. (4) the expressions $\vec{p}_C = \vec{p}_{lab} - m\vec{\upsilon}_A(t)$,

$$\vec{r}_C = \vec{r}_{lab} - \int \vec{\upsilon}_A(t) dt, \quad \cos\Psi_C = \frac{p_{lab} \cos\Psi_{lab} - m_L \upsilon_A(t)}{\sqrt{p_{lab}^2 - 2 p_{lab} m_L \upsilon_A(t) \cos\Psi_{lab} + m_L^2 \upsilon_A^2(t)}}$$

As a result we obtain

$$I_{st} = \frac{n_{L,lab} n_{A,lab} \sigma_{AL}}{(2\pi m_L kT_L)^{\frac{3}{2}}} \exp\left(-\frac{p_{L,lab}^2 - 2m_L p_{L,lab} \upsilon_A(t) \cos\Psi_{lab} + 2m_L^2 \upsilon_A^2(t)}{2m_L kT_L}\right) \cdot$$
$$\cdot \left[\frac{kT_L}{m_L \upsilon_A(t)} \sinh\left(\frac{\upsilon_A(t) \sqrt{p_{L,lab}^2 - 2m_L p_{L,lab} \upsilon_A(t) \cos\Psi_{lab} + m_L^2 \upsilon_A^2(t)}}{kT_L}\right) - \right.$$
$$\left. - \frac{\sqrt{p_{L,lab}^2 - 2m_L p_{L,lab} \upsilon_A(t) \cos\Psi_{lab} + m_L^2 \upsilon_A^2(t)}}{m_L} \cdot \right. \quad (5)$$
$$\left. \cdot \exp\left(-\frac{\upsilon_A(t)(p_{L,lab} \cos\Psi_{lab} - m_L \upsilon_A(t))}{kT_L}\right)\right]$$

From here on we work in the laboratory reference frame, therefore the index "lab" in (5) will be omitted.

We estimate $\varphi_1$, determined by the integral of light particle collision with reaction products.

Let us use an asymptotic form of the collision integral (5) at large $\upsilon_A(t)$:

$$I_{st} \sim f_L^{[0]} n_A \sigma_{LA} \upsilon_A(t) \quad (6)$$

Generalization of the expression (6) in case of the interaction with an extended particle group was found due to its consideration as a sequence of flat layers.

$$I_{st,tot} \sim f_L^{[0]} \sigma_{LA} Q \qquad (7)$$

Where Q means flux density of reaction products.

In order to find $\vec{F} = \frac{1}{n_L} \int I_{st}\left(f_L^{[0]}, f_A\right) m_L \left(\vec{v}_L - \vec{v}_{0,L}\right) d\vec{p}_L$, let us consider the collision integral in form (4). Then

$$\vec{F} = -\frac{1}{n_L} \int m_L \vec{v}_L \frac{n_L n_A \sigma_{AL}}{\left(2\pi m_L kT_L\right)^{\frac{3}{2}}} \exp\left(-\frac{p_L^2 + m_L^2 v_A^2(t)}{2m_L kT_L}\right) \cdot \frac{p_L}{m_L} \exp\left(-\frac{p_L v_A(t)\cos\Psi}{kT_L}\right)$$

Computation of this integral when $v_A \gg \sqrt{\frac{kT_L}{m_L}}$ leads us to expression for the force:

$$F = 2\sqrt{2} m_L v_A^2 n_A \sigma_{LA}$$

Proceeding to a long group of reaction products gives us the following equation

$$F = 2\sqrt{2} Q \sigma_{LA} m_L v_A \qquad (8)$$

Now we assess the part with $\varphi_1$. We obtain from the Eq. (3):

$$\int d\vec{p}_{1,L} \left|\vec{V}_L - \vec{V}_{L,1}\right| d\sigma f_L^{[0]}(\vec{V}_L) f_L^{[0]}(\vec{V}_{L,1})\left[\varphi_1(\vec{V}_L') + \varphi_1(\vec{V}_{L,1}') - \varphi_1(\vec{V}_L) - \varphi_1(\vec{V}_{L,1})\right] \leq$$
$$\leq f_L^{[0]} \sigma_{LL} \varphi_{1,\max} \sqrt{\frac{kT_L}{m_L}} \int d\vec{p}_{1,L} f_L^{[0]}(\vec{p}_{L,1}) \sim f_L^{[0]} \sigma_{LL} \varphi_{1,\max} n_L \sqrt{\frac{kT_L}{m_L}} \qquad (9)$$

Thus, we compare (7), (8) and (9):

$$\left|\varphi_{1,\max}\right| \sim \frac{Q}{n_L} \frac{\sigma_{LA}}{\sigma_{LL}} \left(\sqrt{\frac{m_L}{kT_L}} + \frac{m_L v_A}{kT_L}\right)$$

Since ratio $\frac{n_A}{n_L}$ is small, inequality $\varphi_{1,\max} \ll 1$ that guarantees the validity of hydrodynamic approximation can be obtained.

**Analysis of the Minor Interface Perturbations Evolution**

For the purpose of hydrodynamic [4] analysis let us consider the following model system, consisting of three layers of liquid. Constant acceleration directed opposite the z axis acts in the mid region. The linear stage of the evolution of this system is addressed.

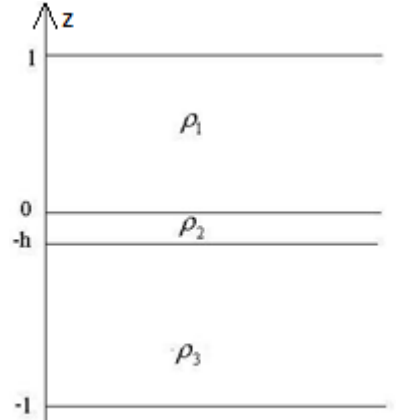

Fig. 1 Three-layer system

Velocity potentials in three regions will satisfy the Laplace's equation

$$\begin{cases} \dfrac{\partial^2 \varphi_1}{\partial x^2} + \dfrac{\partial^2 \varphi_1}{\partial z^2} = 0 \\ \dfrac{\partial^2 \varphi_2}{\partial x^2} + \dfrac{\partial^2 \varphi_2}{\partial z^2} = 0 \\ \dfrac{\partial^2 \varphi_3}{\partial x^2} + \dfrac{\partial^2 \varphi_3}{\partial z^2} = 0 \end{cases} \quad (10)$$

This problem is similar to those discussed in Ref. [4]. In accordance with the above, we specify rigid boundary conditions and seek the solution of the system (10) in the form of potentials of "standing waves" [5]

$$\begin{cases} \varphi_1 = \sin(\omega t)\cos(kx) f_1(z) \\ \varphi_2 = \sin(\omega t)\cos(kx) f_2(z) \\ \varphi_3 = \sin(\omega t)\cos(kx) f_3(z) \end{cases}$$

The equation allows us to determine $f(z)$ in the form:

$$\begin{cases} f_1 = A_1 e^{kz} + B_1 e^{-kz} \\ f_2 = A_2 e^{kz} + B_2 e^{-kz} \\ f_3 = A_3 e^{kz} + B_3 e^{-kz} \end{cases}$$

And then we find $\omega^2$ as an eigen value of the problem.

In general case the expression for $\omega^2$ is rather cumbersome, therefore we will give it with regard for some simplifications (infinitely distant boundaries, $\rho_2 = \rho_3$).

$$\omega^2 = \frac{kg}{4e^{kh}(\rho_1+\rho_2)}\Big(\rho_2 e^{kh} - \rho_2 e^{-kh} - \rho_1 e^{kh} + \rho_1 e^{-kh} \pm$$
$$\pm \sqrt{9\rho_2^2 e^{2kh} - 10\rho_2^2 + 6\rho_1\rho_2 e^{2kh} - 4\rho_1\rho_2 + \rho_2^2 e^{-2kh} - 2\rho_1\rho_2 e^{-2kh} + \rho_1^2 e^{2kh} - 2\rho_1^2 + \rho_1^2 e^{-2kh}}\Big) \quad (11)$$

In the limit $h \ll \lambda$, the expression (11) transfers to:

$$\omega_{1,2}^2 \approx -g\sqrt{\frac{8\pi^3 h}{\lambda^3} \frac{\rho_2}{\rho_2+\rho_1}} \quad (12)$$

At $h \gg \lambda$ the expression (11) will take the form

$$\omega_{3,4}^2 \approx \frac{2\pi g}{4\lambda(\rho_1+\rho_2)}\big(\rho_2 - \rho_1 \pm (3\rho_2 + \rho_1)\big)$$

Accordingly, in this limit two solutions exist:

$$\omega_3^2 = \frac{2\pi g}{\lambda}\frac{\rho_2}{\rho_1+\rho_2} \quad (13)$$

$$\omega_4^2 = -\frac{\pi g}{\lambda} \quad (14)$$

Solution (13) corresponds to frequencies as if vibrations are spreading along the interface of the active and passive media. Solution (14) is the growth increment of small perturbations, spreading along the interface between a passive substance and a layer of reaction products absorption. Thus, because the frequency is imaginary the mixing in the passive medium can develop at any ratio between active and passive media and will manifest itself at the arbitrary length of reaction products deceleration.

We have introduced our MD-results in Fig. 2 and 3. Our referenced system consists of 2400 particles and the constant acceleration acts in inner layer. Our predicted effect was (completely) confirmed: the development of mixing doesn't depend on active/passive media density ratio.

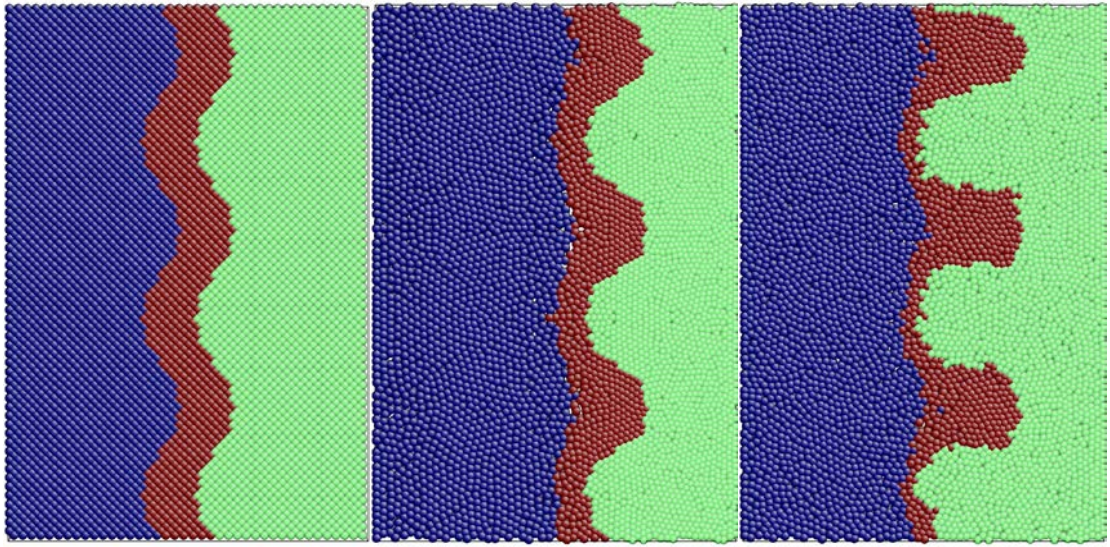

**Fig.2 Instability evolution.**

**Layer density increases from left to right. Acceleration operates in the central layer and is directed from left to right**

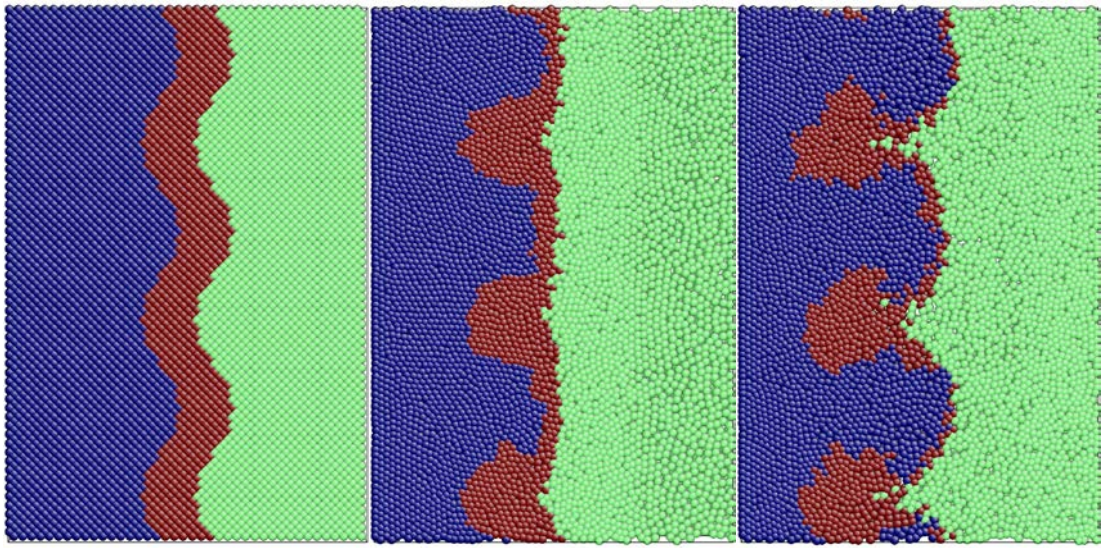

**Fig.3 Instability evolution.**

**Layer density increases from left to right. Acceleration operates in the central layer and is directed from right to left**

## Possible Manifestations of the Instability in Physical Processes

The discussed instability may occur in the following physical processes:

1. Phenomena at the *active- passive media* interface at a thermonuclear explosion or inertial nuclear fusion.
2. Structure of a flame front in supernova Ia.
3. Evolution of hot spots in explosive solids (ES).

The results of the investigation into the hydrodynamic instabilities of media movement with the account of alpha - particle momentum transfer in ICF targets were presented in Ref. [6] The conditions of suppression of the Rayleigh-Taylor and Kelvin-Helmholz instabilities by particle momentum fluxes were obtained.

Unlike Ref. [6], we consider situation, when the whole target has the same density, and classical Rayleigh-Taylor instability doesn't occur.

We assess the growth of a small perturbations. For this purpose we exploit the target characteristics used in NIF [7].

Let's determine the time needed for burning out 50% of tritium. The differential equation for tritium concentration looks as follows:

$$\frac{dn_T}{dt} = -n_D n_T \langle \sigma v \rangle$$

Taking into consideration that $n_T(t) = n_D(t) = n(t)$, we can find the solution of this equation:

$$\frac{n_0 - n}{n_0 n} = t \langle \sigma v \rangle$$

Accordingly, the time during which half the thermonuclear fuel is burnt out is

$$\tau = \frac{1}{n_0 \langle \sigma v \rangle} \qquad (15)$$

The expression for $\langle\sigma\upsilon\rangle$ was taken from Ref. [8] and characterizes the dependence of DT reaction velocity on the temperature.

Based on the data related to the attained maximum density of the DT mixture we estimate the size of the region of compression. If we assume that the target remains spherical till the moment of the maximum compression, we will obtain for the minimum radius $r_{min} = r_0 \sqrt[3]{\frac{\rho_0}{\rho_{max}}}$. According to Ref. [7] we find that $\rho_{min} \approx 7.7\,\mu m$. Next we will address the evolution of a perturbation with a wavelength $\lambda = 5\,\mu m$ and amplitude $a_0 = 1\,\mu m$.

With a perturbation wavelength comparable with the radius of the spherical area, the interface cannot, (undoubtedly), be regarded flat. For instance, if sphericity is taken into account, the exponential law of perturbation growth is replaced by a power law for some types of flame hydrodynamic instability [9]. Nevertheless at this stage we will make assessments for a flat boundary.

In order to understand if it is possible to use approximation (12), it is necessary to estimate a free path of an $\alpha$ - particle.

According to the Ref. [10], a path length of 4.2 MeV $\alpha$-particle in a substance with the density $490\,\frac{g}{cm^3}$ (in accordance with max. density of DT-gas, taken from Ref. [7]) is $2.3 \cdot 10^{-8}\,m$. Thus, the approximation (12) for this problem is applicable and allows to determine a growth increment

$$\omega^2 = -g\sqrt{\frac{8\pi^3 h}{\lambda^3}\frac{\rho_{outer}}{\rho_{outer}+\rho_{inner}}} \quad (16)$$

It is necessary to determine the acceleration given to the outer layers by a flux of $\alpha$-particles. With the help of simple estimations (detailed computations for the similar case in ES are given below) we find that $g = \frac{\langle\sigma\upsilon\rangle}{4\rho_{outer}}n^2\sqrt{2m_{He}E_0}$ and the expression (16) takes the form:

$$\omega^2 = -\langle\sigma\upsilon\rangle n^2 \sqrt{\frac{m_{He}E_0\pi^3 h}{\lambda^3 \rho_{outer}(\rho_{outer}+\rho_{inner})}} \qquad (17)$$

As at the linear stage $A(t) \approx a_0 e^{\omega t}$, based on (15) and (17), we can write the explicit form for $\omega\tau$:

$$\omega\tau = \frac{1}{\sqrt{\langle\sigma\upsilon\rangle}}\left[\frac{m_{He}E_0\pi^3 h}{\lambda^3 \rho_{outer}(\rho_{outer}+\rho_{inner})}\right]^{\frac{1}{4}} \qquad (18)$$

Then we use the parameters from Ref. [7,8] and find that $\omega\tau \approx 7$, i.e., the effect of mixing of DT-mixture outer layers caused by the development of the instability can be regarded significant.

Let us consider the behavior (18) as the functions of the maximum temperature and density. The speed of DT thermonuclear reaction does not practically change with further temperature growth. If the density increases by $n$ times, the path length of an $\alpha$ particle will become $\frac{h}{n}$, and the perturbation wavelength - $\frac{\lambda}{n^{\frac{1}{3}}}$, therefore an index of the exponent will finally have the value $\frac{(\omega\tau)_0}{\sqrt[4]{n}}$. This result gives some grounds to confirm the increase of power provided to the target in the existing and projected facilities of laser thermonuclear fusion.

Now we consider possible influence of this instability on the structure of a flame front in SNIa.

According to SNIa models, an explosion begins with deflagration of $C_6^{12}$. We consider the energy of an $\alpha$ − particle generated as a result of the following reaction:

$C_6^{12} + C_6^{12} \to Mg_{12}^{24} \to Ne_{10}^{20} + He_2^4 + \gamma$. For the above (consideration) let us consider the case without photon emission. Then the energy of produced particles can be determined accurately.

$$\varepsilon_{He_2^4} = \frac{\left(2M_{C_6^{12}}c^2\right)^2 + \left(M_{He_2^4}c^2\right)^2 - \left(M_{Ne_{10}^{20}}c^2\right)^2}{2\cdot 2M_{C_6^{12}}c^2}, \text{ where } \varepsilon_{He_2^4} = M_{He_2^4}c^2 + E_{He_2^4}$$

Thus, $E_{He_2^4} \approx 4 MeV$

If $\rho = 10^7 \frac{g}{cm^3}$, according to Ref. [10], the path length of an $\alpha$-particle is $3.5 \cdot 10^{-12} m$. At the same time, according to Ref. [11], the width of a flame front exceeds $5 \cdot 10^{-8} m$ even if the density is much bigger. Therefore this type of instability cannot effect the development large-scale perturbations of the flame front.

We consider as an example the medium with $\rho = 2 \cdot 10^9 \frac{g}{cm^3}$ and perturbation with a wavelength that is approximately equal to the width of a flame front. According to (12) $\omega \sim 10^{15} s^{-1}$.

The inverse value of this frequency can be to some extend equated to the time of substance turbulization inside the burning front.

Finally, let us consider the influence of discussed instability in ES-detonation processes.

Nowadays models of ES detonation are developed intensively. They are all based on the model of hot-spots (HS), namely, local heating areas of an explosive substance subjected to a shock wave. However the description of the process of their evolution, with the hydrodynamics and Arrhenius kinetics, hot spots do not merge in their increase and decay.

Thus a problem related to the mechanism of a hot spot evolution in solid heterogeneous explosives arises.

The stage of HS formation lasts approximately 1 ns [12], then hot spots grow until they get into contact. The overall time of the whole process is much greater and determines the detonation induction ~ 1 μs. If we mean that average size of explosive granules are ~ 10÷100 μm, the burning speed must be of the order of 100 m/s.

According to Ref. [9], turbulization of a flame front in gases is an effective factor that increases the burning speed. Thus let us consider a contribution of the discussed instability in the process of the development of small perturbations at the interface *explosive substance – explosion products*.

Suppose that N fragments of ES molecules are produced in inner spherical layer as a result of thermal decomposition. By virtue of isotropy half of them entered in outer spherical layer 2. Averaging by a flight-in angle, we suppose that this is equivalent to one-fourth of original fragments flying strictly along the radiuses.

Based on the results of molecular-dynamic calculations it was determined that the free path of ES molecular fragments is equal to 2-3 intermolecular distances. Therefore each fragment flying into the passive medium may be imagined as completely losing its momentum already in the second layer. So, the transmitted momentum can be described by the following equation:

$$\frac{N}{4}\sqrt{2m_{fragment}E_{fragment}} = M_{outer}V_{outer}$$

Having differentiated this expression in time, we obtain the following

$$\frac{Q}{4}\sqrt{2m_{fragment}E_{fragment}} = M_{outer}g$$

The mass of a thin spherical layer $M_{outer} = 4\pi r^2 \cdot h \cdot \rho_{ES}$

$$Q = \frac{N_{fragment}}{\Delta t}, \text{ where } \Delta t = \frac{h}{\upsilon_{combustion}}$$

According to Ref. [13], for PETN $N_{fragment} = 4N_{PETN}$. Then we will carry out the estimation for this explosive substance.

After rather simple conversions we obtain

$$N_{PETN} = \frac{M_{inner}}{Mr_{PETN}}N_A = \frac{4\pi r^2 \cdot h \cdot \tilde{\rho}_{PETN}}{Mr_{PETN}}N_A$$

$$Q = \frac{16\pi\tilde{\rho}_{PETN}\upsilon_{combustion}r^2}{Mr_{PETN}}N_A$$

Thus, $g = \dfrac{N_A \cdot \upsilon_{combustion}}{Mr_{PETN} \cdot h}\sqrt{2m_{fragment}E_{fragment}}$  (19)

Substituting (26) in (23), we finally find that

$$\omega^2 = -\frac{4\upsilon_{combustion}N_A}{Mr_{PETN}}\sqrt{\frac{\pi^3 m_{fragment}E_{fragment}}{\lambda^3 h}\frac{\rho_{ES}}{\rho_{ES}+\rho_{EP}}} \qquad (20)$$

Based on the results by Ref. [13], we find from (20) that for perturbation with $\lambda = 4\mu m$, $\omega\tau \sim 260$, which can testify to a significant contribution of the discussed instability in the turbulization process of hot spot boundary burning.

**Conclusion**

The paper deals with estimations of the correction for the equilibrium function of passive medium particle distribution. Based on these estimations the limitations of the applicability of hydrodynamic approximation were determined. Using the potential flow theory, a growth increment of small perturbations of the interface of the active and passive media is found.

We consider emergence and possible consequences of instability development for several physical processes studied in our paper: phenomena at the *active- passive media* interface at a thermonuclear explosion or inertial nuclear fusion, structure of a flame front in SNIa, evolution of hot spots in ES. For typical time periods, we observed in each case substantial increases in initial perturbance which go beyond applicability of linear approximation. This can indicate the contribution of this instability to the process of turbulence in the boundary line is significant.